\documentclass[aps,graphicx,9pt,showkeys,showpacs,twocolumn]{revtex4}
\usepackage{amsmath}
\usepackage{amscd}
\usepackage{graphicx}
\usepackage{subfigure}
\usepackage{appendix}
\begin{document}

\title[Short Title]{Invariant-based pulse design for three-level systems without the rotating-wave approximation}

\author{Yi-Hao Kang$^{1,2}$}
\author{Ye-Hong Chen$^{1,2}$}
\author{Bi-Hua Huang$^{1,2}$}
\author{Jie Song$^{3}$}
\author{Yan Xia$^{1,2,}$\footnote{E-mail: xia-208@163.com}}

\affiliation{$^{1}$Department of Physics, Fuzhou University, Fuzhou 350002, China\\
             $^2$Fujian Key Laboratory of Quantum Information and Quantum Optics (Fuzhou University), Fuzhou 350116, China\\
             $^{3}$Department of Physics, Harbin Institute of Technology, Harbin 150001, China}

\begin{abstract}
In this paper, a scheme is put forward to design pulses which drive
a three-level system based on the reverse engineering with
Lewis-Riesenfeld invariant theory. The scheme can be applied to a
three-level system even when the rotating-wave approximation (RWA)
can not be used. The amplitudes of pulses and the maximal values of
detunings in the system could be easily controlled by adjusting
control parameters. We analyze the dynamics of the system by an
invariant operator, so additional couplings are unnecessary.
Moreover, the approaches to avoid singularity of pulses are studied
and several useful results are obtained. We hope the scheme could
contribute to fast quantum information processing without RWA.

\pacs{03.67. Pp, 03.67. Mn, 03.67. HK}

\keywords{Shortcut to adiabatic passage; Lewis-Riesenfeld invariant;
Three-level system; Rotating-wave approximation}
\end{abstract}

\maketitle

\section{INTRODUCTION}

Manipulating physical systems with time-dependent electromagnetic
fields, which is important for high-precision quantum information
processing, has attracted growing interests in recent years. The
adiabatic passage
\cite{FewellAJP50,BergmannRMP70,VitanovARPC52,KralRMP79} is one of
typical methods to design and control time-dependent pulses, which
has been widely used in numerous previous schemes
\cite{ZhengPRL95,YangPRL92,MolerPRA75,DengPRA74}. The adiabatic
passage is approved for its robustness against the fluctuations of
control parameters, while it is also criticized for the low speed
caused by the limit of adiabatic condition. To accelerate evolutions
of physical systems, many methods
\cite{DemirplakJPCA107,BerryJPA42,ChenPRL105,CampoPRL111,CampoPRL109,CampoPRA84,CampoEPL96,CampoSR2,DeffnerPRX4,ChenPRA82,ChenPRA83,ChenPRA84,MugaJPB42,MugaJPB43,ChenPRL104}
have been proposed. Since they are related to the adiabatic passage,
but provide alternative paths without the adiabatic condition for
evolutions of physical systems, these methods
\cite{DemirplakJPCA107,BerryJPA42,ChenPRL105,CampoPRL111,CampoPRL109,CampoPRA84,CampoEPL96,CampoSR2,DeffnerPRX4,ChenPRA82,ChenPRA83,ChenPRA84,MugaJPB42,MugaJPB43,ChenPRL104}
are arranged as a new kind of technique named by shortcuts to
adiabaticity (STA). In the past several years, STA has drawn much
attention of researchers, and has subsequently been used in many
physical systems, such as superconducting systems
\cite{KangPRA94,ZhangSR5}, atom-cavity systems
\cite{ChenPRA86,yehongPRA89}, and spin-NV center systems
\cite{SongPRA93,SongNJP18}. Besides, many schemes
\cite{IbanezPRL109,IbanezPRA87,IbanezPRA89,yehongPRA93,TorronteguiPRA89,GaraotPRA89,qichengOE24,KangSR6}
have been put forward to improve or extend STA. Until now, STA could
be used to design pulses perfectly in many different cases.

The previous schemes
\cite{DemirplakJPCA107,BerryJPA42,ChenPRL105,CampoPRL111,CampoPRL109,CampoPRA84,CampoEPL96,CampoSR2,
DeffnerPRX4,ChenPRA82,ChenPRA83,ChenPRA84,MugaJPB42,MugaJPB43,ChenPRL104,KangPRA94,ZhangSR5,ChenPRA86,
yehongPRA89,SongPRA93,SongNJP18,IbanezPRL109,IbanezPRA87,IbanezPRA89,yehongPRA93,TorronteguiPRA89,GaraotPRA89,qichengOE24,KangSR6}
with STA focused on the physical systems under the rotating-wave
approximation (RWA). However, many recent schemes about
superconducting systems \cite{SornborgerPRA70,LiuPRA90,SankPRL117},
optomechanical systems \cite{MalzPRA94}, semiconducting systems
\cite{SongPRA94}, Bose-Einstein condensates \cite{HofferberthPRA76},
and NV centers \cite{ScheuerNJP16} have shown that, RWA may be
invalid in the cases of ultra-fast operations and ultra-strong
couplings. For example, Liu \emph{et al.} \cite{LiuPRA90} have shown
that RWA is broken down in the ultra-strong coupling, where the
frequencies of pulses take the value of $10\times2\pi$GHz and the
coupling strengthes take the value of $1.021\times2\pi$GHz.
Moreover, Scheuer \emph{et al.} \cite{ScheuerNJP16} have
demonstrated in a NV center that, when using a magnetic field with a
frequency of 30MHz, RWA can not be used for a qubit control if the
Rabi frequency larger than 15MHz. From these examples
\cite{SornborgerPRA70,LiuPRA90,SankPRL117,MalzPRA94,SongPRA94,HofferberthPRA76,ScheuerNJP16},
RWA may be invalid in fast quantum information processing, thus the
applications of previous schemes
\cite{DemirplakJPCA107,BerryJPA42,ChenPRL105,CampoPRL111,CampoPRL109,CampoPRA84,CampoEPL96,CampoSR2,
DeffnerPRX4,ChenPRA82,ChenPRA83,ChenPRA84,MugaJPB42,MugaJPB43,ChenPRL104,KangPRA94,ZhangSR5,ChenPRA86,
yehongPRA89,SongPRA93,SongNJP18,IbanezPRL109,IbanezPRA87,IbanezPRA89,yehongPRA93,TorronteguiPRA89,GaraotPRA89,qichengOE24,KangSR6}
with STA would be limited. Therefore, it is worthwhile to study STA
without RWA so that pulse design for fast quantum information
processing could be more effective.

Last year, two schemes \cite{IbanezPRA92,ChenPRA91} have been
proposed, which are about STA without RWA. One scheme
\cite{ChenPRA91} is proposed by Chen \emph{et al.}, in which
transitionless quantum driving (the counterdiabatic driving) is
exploited to investigate the dynamics of both two- and three-level
systems. It has shown that population transfers in both two- and
three-level systems could be achieved in theory. The scheme
\cite{ChenPRA91} is interesting, but it has a few disadvantages.
First, using the transitionless quantum driving requires an extra
coupling between the initial state and the final state, which may be
hard to be realized in several cases. Besides, the control
parameters are not flexible enough to control the amplitudes of
pulses and maximal values of detunings. Moreover, how to reduce
oscillations and avoid the singularity of pulses, which are two
questions required to be considered when RWA is broken down, have
not been discussed. The other interesting scheme \cite{IbanezPRA92}
is proposed by Ib\'{a}\~{n}ez \emph{et al.}, which is about pulse
design for a two-level system with both transitionless quantum
driving and invariant-based method with Lewis-Riesenfeld theory
\cite{LewisJMP10}. Their scheme \cite{IbanezPRA92} has shown many
interesting results. For example, using invariant-based method does
not require any extra couplings, which makes the pulse design more
feasible in experiments. Moreover, the singularity of pulses can be
avoided by choosing control parameters suitably. Furthermore, they
have shown that an invariant-based pulse design can help to achieve
a population transfer in a two-level system with a perfect fidelity.
These interesting results have demonstrated that the invariant-based
method is very promising. However, different systems possess
different dynamic features. Invariants for a two-level system
without RWA can not properly describe the dynamics of a three-level
system without RWA. Moreover, with the dimensions increase, the
complexity of invariants would greatly increase. Therefore, the
scheme \cite{IbanezPRA92} can not be directly applied to a
three-level system without RWA. But three-level systems are very
important in quantum information processing, as many quantum
information tasks can be implemented in physical systems which are
equivalent or approximately equivalent to three-level systems
\cite{lumeiPRA89,yehongPRA91,LonghiLPR3,LonghiJPB44,OrnigottiJPB41,RangelovPRA85}.
So it is necessary to research dynamics of three-level systems
without RWA. Considering the advantages of the invariant-based
method, if it can be applied to pulse design for three-level systems
without RWA, we can realize many interesting quantum information
tasks with ultra-fast operations and ultra-strong couplings. That
requires us firstly to find out an invariant for three-level systems
without RWA.

In this paper, inspired by the schemes \cite{IbanezPRA92,ChenPRA91},
we propose a scheme to design pulses for a three-level system
without RWA. The scheme is based on a new-found invariant operator,
which can help to study the dynamics of a three-level system without
RWA. The scheme has some advantages, such as high speed, robustness
against fluctuations of parameters, no requirements on extra
couplings, etc.. These advantages would be clearly shown in the
following sections.

The article is organized as follows. In Sec. II, we briefly review
the Lewis-Riesenfeld invariant theory. In Sec. III, we give an
invariant for a three-level system without RWA. Based on this
invariant, the mathematical expressions of pulses and detunings are
determined. In Sec. IV, we complete population transfers for a
three-level system without RWA as examples to show the validity of
the scheme. Finally, the conclusions are given in Sec. V.

\section{Lewis-Riesenfeld invariant theory}

In this section, let us briefly introduce Lewis-Riesenfeld theory
\cite{LewisJMP10}. We consider a quantum system which has a
time-dependent Hamiltonian $H(t)$. Now, we introduce an invariant
Hermitian operator $I(t)$, which satisfies ($\hbar=1$)
\begin{equation}\label{lr1}
i\frac{\partial}{\partial t}I(t)-[H(t),I(t)]=0.
\end{equation}
If $|\psi(t)\rangle$ is a solution of the time-dependent
Schr\"{o}dinger equation
$i\partial_{t}|\psi(t)\rangle=H(t)|\psi(t)\rangle$,
$I(t)|\psi(t)\rangle$ is a solution as well. Moreover,
$|\psi(t)\rangle$ can be expanded by eigenvectors of $I(t)$ as
\begin{equation}\label{lr2}
|\psi(t)\rangle=\sum\limits_{k}C_ke^{i\theta_k}|\phi_k(t)\rangle,
\end{equation}
where, $|\phi_k(t)\rangle$ is the $k$th eigenvector of $I(t)$, and
$C_k=\langle\phi_k(0)|\psi(0)\rangle$ is the corresponding
coefficient. $\theta_k$ is the Lewis-Riesenfeld phase for
$|\phi_k(t)\rangle$, which satisfies
\begin{equation}\label{lr3}
\dot{\theta}_k=\langle\phi_k(t)|i\partial_{t}-H(t)|\phi_k(t)\rangle,
\end{equation}
with $\theta_k(t_i)=0$ ($t_i$ is the initial time).

\section{Invariant-based pulse design for three-level systems without the rotating-wave approximation}

Let us start with a three-level system with two ground states
$|1\rangle$, $|3\rangle$, and an excited state $|2\rangle$ shown in
Fig. 1.
\begin{figure}
\scalebox{0.8}{\includegraphics[scale=1]{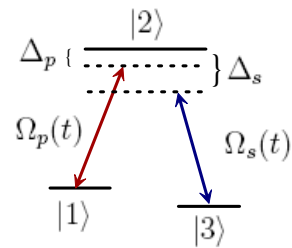}} \caption{The
energy levels of the three-level system.}
\end{figure}
Without RWA, the Hamiltonian of this system can be written by
\begin{eqnarray}\label{e1}
H(t)=\left[%
\begin{array}{ccc}
  -\omega_p-\Delta_p(t) &\  \Omega_p(t)\cos(\omega_pt) &\  0 \\
 \Omega_p^*(t)\cos(\omega_pt) &\  0 &\  \Omega_s^*(t)\cos(\omega_st) \\
  0 &\  \Omega_s(t)\cos(\omega_st) &\  -\omega_s-\Delta_s(t) \\
\end{array}%
\right],
\end{eqnarray}
in basis $\{|1\rangle,|2\rangle,|3\rangle\}$, where, $\Omega_p(t)$,
$\Omega_s(t)$ are the pump and Stokes pulses driving the transitions
$|1\rangle\leftrightarrow|2\rangle$ and
$|3\rangle\leftrightarrow|2\rangle$, respectively. $\Omega_p^*(t)$,
$\Omega_s^*(t)$ are the complex conjugates of $\Omega_p(t)$,
$\Omega_s(t)$, respectively. $\omega_p$ and $\omega_s$ are the
frequencies of pump and Stokes pulses, respectively. $\Delta_p(t)$
and $\Delta_s(t)$ denote the detunings of the pump and Stokes pulses
from their relevant transitions, respectively. By analyzing
Eqs.~(\ref{lr1}) and (\ref{e1}) with some undetermined coefficients,
we find out a Lewis-Riesenfeld invariant as follows:
\begin{eqnarray}\label{e2}
I(t)=\left[%
\begin{array}{ccc}
  I_{11} &\  I_{12} &\  I_{13} \\
 I_{12}^* &\  I_{22} &\  I_{23} \\
  I_{13}^* &\  I_{23}^* &\  I_{33}\\
\end{array}%
\right],
\end{eqnarray}
and the matrix elements of $I(t)$ are given as
\begin{eqnarray}\label{e3}
&I_{11}&=\cos2\lambda(\cos^2\alpha\cos^2\beta-\sin^2\alpha)\cr\cr&&+\cos\epsilon\cos\beta\sin2\alpha\sin2\lambda,\cr\cr
&I_{12}&=(\cos\alpha\cos2\lambda\cos\beta+e^{-i\epsilon}\sin\alpha\sin2\lambda)\sin\beta,\cr\cr
&I_{13}&=\frac{1}{4}\cos2\lambda(3+\cos2\beta)\sin2\alpha\cr\cr&&-\cos\beta(\cos\epsilon\cos2\alpha+i\sin\epsilon)\sin2\lambda,\cr\cr
&I_{22}&=\cos2\lambda\sin^2\beta,\cr\cr
&I_{23}&=(\sin\alpha\cos2\lambda\cos\beta-e^{i\epsilon}\cos\alpha\sin2\lambda)\sin\beta,\cr\cr
&I_{33}&=\cos2\lambda(\sin^2\alpha\cos^2\beta-\cos^2\alpha)\cr\cr&&-\cos\epsilon\cos\beta\sin2\alpha\sin2\lambda.
\end{eqnarray}
In Eq.~(\ref{e3}), $\alpha$, $\beta$, $\epsilon$, $\lambda$ are four
auxiliary time-dependent parameters, and they are required to
satisfy
\begin{eqnarray}\label{e4}
\dot{\alpha}=\dot{\lambda}\cos\beta\cos\epsilon.
\end{eqnarray}
The invariant $I(t)$ has three eigenvectors as follows:
\begin{eqnarray}\label{e5}
&&
|\phi_{+}(t)\rangle=\left[%
\begin{array}{c}
  \cos\alpha\cos\beta\cos\lambda+e^{i\epsilon}\sin\alpha\sin\lambda \\
  \sin\beta\cos\lambda \\
  \sin\alpha\cos\beta\cos\lambda-e^{i\epsilon}\cos\alpha\sin\lambda \\
\end{array}%
\right],\cr\cr&&
|\phi_{-}(t)\rangle=\left[%
\begin{array}{c}
  \cos\alpha\cos\beta\sin\lambda-e^{i\epsilon}\sin\alpha\cos\lambda \\
  \sin\beta\sin\lambda \\
  \sin\alpha\cos\beta\sin\lambda+e^{i\epsilon}\cos\alpha\cos\lambda \\
\end{array}%
\right],\cr\cr&&
|\phi_0(t)\rangle=\left[%
\begin{array}{c}
  \cos\alpha\sin\beta \\
  -\cos\beta \\
  \sin\alpha\sin\beta \\
\end{array}%
\right],
\end{eqnarray}
which corresponds to eigenvalues 1, -1 and 0 of $I(t)$.

Solving Eq.~(\ref{lr1}) with $H(t)$ in Eq.~(\ref{e1}) and $I(t)$ in
Eq.~(\ref{e2}), we obtain the following results
\begin{eqnarray}\label{e65}
&\Omega_p(t)\cos(\omega_pt)&=i\dot{\lambda}e^{-i\epsilon}\sin\alpha\sin\beta\cr\cr&&+\frac{1}{2}\cos\alpha(-2i\dot{\beta}+\dot{\theta}\sin2\beta),\cr\cr
&\Omega_s(t)\cos(\omega_st)&=-i\dot{\lambda}e^{-i\epsilon}\cos\alpha\sin\beta\cr\cr&&+\frac{1}{2}\sin\alpha(-2i\dot{\beta}+\dot{\theta}\sin2\beta),\cr\cr
&\omega_p+\Delta_p(t)&=-\dot{\epsilon}\sin^2\alpha+\dot{\theta}(\cos^2\alpha\sin^2\beta-\cos^2\beta)\cr\cr&&-\dot{\lambda}\sin\epsilon\sin2\alpha\cos\beta,\cr\cr
&\omega_s+\Delta_s(t)&=-\dot{\epsilon}\cos^2\alpha+\dot{\theta}(\sin^2\alpha\sin^2\beta-\cos^2\beta)\cr\cr&&+\dot{\lambda}\sin\epsilon\sin2\alpha\cos\beta.
\end{eqnarray}
In Eq~(\ref{e65}), $\theta$ is the Lewis-Riesenfeld phase of
$|\phi_0(t)\rangle$, which could be solved by
\begin{equation}\label{e75}
\dot{\theta}=\langle\phi_0(t)|i\partial_{t}-H(t)|\phi_0(t)\rangle=-\frac{\dot{\epsilon}+2\dot{\lambda}\sin\epsilon\cos\beta\cot2\alpha}{\sin^2\beta}.
\end{equation}
Besides, the Lewis-Riesenfeld phases of $|\phi_+(t)\rangle$ and
$|\phi_-(t)\rangle$ are both zero.

With the results above, we can use the following formula to
calculate the evolution of the system
\begin{eqnarray}\label{e85}
&|\psi(t)\rangle&=(\langle\phi_+(0)|\psi(0)\rangle)|\phi_+(t)\rangle+(\langle\phi_-(0)|\psi(0)\rangle)|\phi_-(t)\rangle\cr\cr&&+e^{i\theta}(\langle\phi_0(0)|\psi(0)\rangle)|\phi_0(t)\rangle.
\end{eqnarray}

\section{Population transfers for a three-level system}

Using the results shown in Sec. III, we would like to perform
population transfers for a three-level system to check the validity
of the scheme. For simplicity, the condition
$\dot{\lambda}=\dot{\alpha}=0$ is set, so
$\dot{\epsilon}=-\dot{\theta}\sin^2\beta$ can be obtained from
Eq.~(\ref{e75}), and Eq.~(\ref{e65}) reduces to
\begin{eqnarray}\label{e6}
&\Omega_p(t)\cos(\omega_pt)&=\frac{1}{2}\cos\alpha(-2i\dot{\beta}+\dot{\theta}\sin2\beta),\cr\cr
&\Omega_s(t)\cos(\omega_st)&=\frac{1}{2}\sin\alpha(-2i\dot{\beta}+\dot{\theta}\sin2\beta),\cr\cr
&\omega_p+\Delta_p(t)&=-\dot{\theta}\cos2\beta,\cr\cr
&\omega_s+\Delta_s(t)&=-\dot{\theta}\cos2\beta.
\end{eqnarray}
Then, two time-independent coefficients $\kappa_p$ and $\kappa_s$
are introduced, such that
\begin{eqnarray}\label{e7}
&\omega_p+\Delta_p(t)&=-\kappa_p\dot{\theta}+(\kappa_p-\cos2\beta)\dot{\theta},\cr\cr
&\omega_s+\Delta_s(t)&=-\kappa_s\dot{\theta}+(\kappa_s-\cos2\beta)\dot{\theta}.
\end{eqnarray}
In addition, we introduce a positive time-independent parameter
$\omega$, which has the scale of frequency. Assuming
$\dot{\theta}=-\omega$, Eq.~(\ref{e7}) can be replaced by
\begin{eqnarray}\label{e8}
&\omega_p+\Delta_p(t)&=\kappa_p\omega-(\kappa_p-\cos2\beta)\omega,\cr\cr
&\omega_s+\Delta_s(t)&=\kappa_s\omega-(\kappa_s-\cos2\beta)\omega.
\end{eqnarray}
Furthermore, Eq.~(\ref{e8}) can be rewritten by
\begin{eqnarray}\label{e9}
&\omega_p&=\kappa_p\omega,\cr\cr &\omega_s&=\kappa_s\omega,\cr\cr
&\Delta_p(t)&=-(\kappa_p-\cos2\beta)\omega,\cr\cr
&\Delta_s(t)&=-(\kappa_s-\cos2\beta)\omega.
\end{eqnarray}
For a brief discussion, we consider that the pump and Stokes pulses
have the same frequency $\omega_p=\omega_s$, but different
polarization directions. Besides, the two-photon resonance
condition, where $\Delta_p(t)=\Delta_s(t)=\Delta(t)$, is considered.
With assumptions shown above, a simple choice is to set
$\kappa_p=\kappa_s=1$, such that
\begin{eqnarray}\label{e10}
&\omega_p=\omega_s&=\omega,\cr\cr
&\Delta(t)&=-2\omega\sin^2\beta,\cr\cr &\Omega_p(t)\cos(\omega
t)&=-\frac{\cos\alpha}{2}(2i\dot{\beta}+\omega\sin2\beta),\cr\cr
&\Omega_s(t)\cos(\omega
t)&=-\frac{\sin\alpha}{2}(2i\dot{\beta}+\omega\sin2\beta).
\end{eqnarray}
In the following, we design the parameters from different viewpoints
and analyze the physical feasibility of the population transfers.

\subsection{Pulse design with smooth functions}

We suppose that a population transfer starts at $t=0$ and ends at
$t=T$. And the initial state of the system is
$|\psi(0)\rangle=|1\rangle$. Considering the following requirements:

(i) The pump and Stokes pulses could be smoothly turned on and
turned off.

(ii) To avoid the singularity of the pump and Rabi frequencies of
Stokes pulses.

(iii) To avoid overlarge detunings or pulses.

$\beta$ and its time derivative $\dot{\beta}$ can be designed as
follows:
\begin{eqnarray}\label{e11}
&\beta&=\frac{A}{2}[1-\cos(\frac{2\pi t}{T})]\cos^2(\omega t),\cr\cr
&\dot{\beta}&=\frac{\pi A}{T}\sin(\frac{2\pi t}{T})\cos^2(\omega
t)\cr\cr&&-\frac{A\omega}{2}[1-\cos(\frac{2\pi t}{T})]\sin(2\omega
t),
\end{eqnarray}
where $A$ is a time-independent coefficient controlling the maximal
value of $\beta$. From Eq.~(\ref{e11}), we have
\begin{eqnarray}\label{e12}
\beta(0)=\beta(T)=\dot{\beta}(0)=\dot{\beta}(T)=0,
\end{eqnarray}
so the pump and Stokes pulses could be smoothly turned on and turned
off. Moreover, when $A$ is not too large, the maximal values of
detunings and amplitudes of pulses could be controlled in desired
ranges. Besides, substituting Eq.~(\ref{e11}) into Eq.~(\ref{e10}),
one can find that when $\cos(\omega t)\rightarrow0$, we have
$\sin2\beta/\cos(\omega t)\rightarrow0$ and $\dot{\beta}/\cos(\omega
t)\rightarrow \mathrm{Const}$. Therefore, the singularity of Rabi
frequencies of the pump and Stokes pulses can be eliminated.

On the other hand, using Eqs.~(\ref{e85}) and (\ref{e12}), the final
state of the system can be obtained
\begin{eqnarray}\label{e13}
|\psi(T)\rangle=\left[%
\begin{array}{c}
  \cos^2\alpha+e^{i\epsilon}\sin^2\alpha \\
  0 \\
  (1-e^{i\epsilon})\sin\alpha\cos\alpha \\
\end{array}%
\right].
\end{eqnarray}
By choosing $\alpha=\pi/4$, we have $|\psi(T)\rangle=|3\rangle$ when
\begin{eqnarray}\label{e14}
\epsilon(T)=\omega\int_{0}^{T}\sin^2\beta dt=\pi.
\end{eqnarray}

Solving Eq.~(\ref{e14}) with numerical methods, some samples of the
relations between $A$ and $\omega T$ are given in Table I.
\begin{center}
\centering{\bf Table I. $A$ with corresponding $\omega T$.}
{\small\begin{tabular}{cc} \hline\hline
\ \ \ \ \ \ \ \ \ \ \ $A$\ \ \ \ \ \ \ \ \ \ \ \ \ \ \ \ \ \ \ \ \ \ &\ \ \ \ \ \ \ \ \ \ \ \ \ \ \ \ \ \ \ \ \ \ \ \ \ $\omega T$\ \ \ \ \ \ \ \ \\
\hline
\ \ \ \ \ \ \ \ \ \ \ $0.2$\ \ \ \ \ \ \ \ \ \ \ \ \ \ \ \ \ \ \ \ \ \ &\ \ \ \ \ \ \ \ \ \ \ \ \ \ \ \ \ \ \ \ \ \ \ \ $179.04\pi$\ \ \ \ \ \ \ \ \ \\
\ \ \ \ \ \ \ \ \ \ \ $0.3$\ \ \ \ \ \ \ \ \ \ \ \ \ \ \ \ \ \ \ \ \ \ &\ \ \ \ \ \ \ \ \ \ \ \ \ \ \ \ \ \ \ \ \ \ \ \ $80.28\pi$\ \ \ \ \ \ \ \ \ \\
\ \ \ \ \ \ \ \ \ \ \ $0.4$\ \ \ \ \ \ \ \ \ \ \ \ \ \ \ \ \ \ \ \ \ \ &\ \ \ \ \ \ \ \ \ \ \ \ \ \ \ \ \ \ \ \ \ \ \ \ $45.72\pi$\ \ \ \ \ \ \ \ \ \\
\ \ \ \ \ \ \ \ \ \ \ $0.5$\ \ \ \ \ \ \ \ \ \ \ \ \ \ \ \ \ \ \ \ \ \ &\ \ \ \ \ \ \ \ \ \ \ \ \ \ \ \ \ \ \ \ \ \ \ \ $29.73\pi$\ \ \ \ \ \ \ \ \ \\
\ \ \ \ \ \ \ \ \ \ \ $0.6$\ \ \ \ \ \ \ \ \ \ \ \ \ \ \ \ \ \ \ \ \ \ &\ \ \ \ \ \ \ \ \ \ \ \ \ \ \ \ \ \ \ \ \ \ \ \ $21.05\pi$\ \ \ \ \ \ \ \ \ \\
\ \ \ \ \ \ \ \ \ \ \ $0.7$\ \ \ \ \ \ \ \ \ \ \ \ \ \ \ \ \ \ \ \ \ \ &\ \ \ \ \ \ \ \ \ \ \ \ \ \ \ \ \ \ \ \ \ \ \ \ $15.83\pi$\ \ \ \ \ \ \ \ \ \\
\hline \hline
\end{tabular}}
\end{center}

When $A=0.2$ ($A=0.3$), a population transfer would go through about
90 (40) pulse periods, which makes the Rabi frequencies of pump and
Stokes pulses oscillate very quickly. Therefore, we focus on the
cases when $A=0.4,0.5,0.6,0.7$ in the following. We define the
population of state $|j\rangle$ as $P_j(t)=|\langle
j|\psi(t)\rangle|^2$ ($j=1,2,3$). In addition, since $\alpha=\pi/4$
is chosen, we have $\Omega_s(t)=\Omega_p(t)=\Omega(t)$. The
populations $P_1$, $P_2$, and $P_3$ versus $t/T$ with different
parameters are shown in Fig. 2. Besides, the real (imaginary) part
$\mathrm{Re}[\Omega(t)]$ ($\mathrm{Im}[\Omega(t)]$) of $\Omega(t)$
and the detuning $\Delta(t)$ versus $t/T$ with different parameters
are shown in Fig. 3.

\begin{figure}
\scalebox{0.45}{\includegraphics[scale=1]{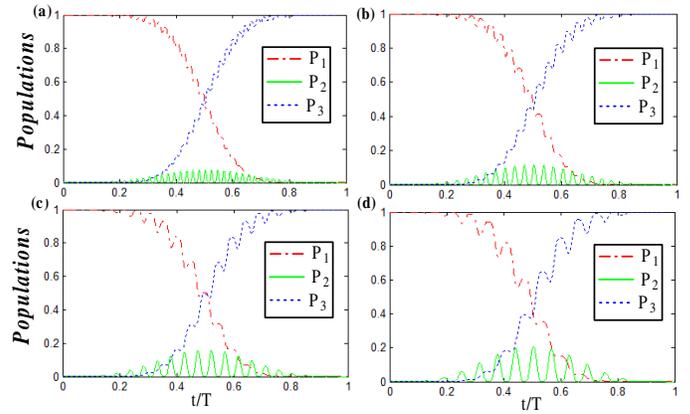}}
\caption{Populations $P_1$ (the dashed-dotted red line), $P_2$ (the
solid green line), and $P_3$ (the dotted blue) versus $t/T$ with
different parameters: (a) $A=0.4$, $\omega=45.7220\pi/T$; (b)
$A=0.5$, $\omega=29.7323\pi/T$; (c) $A=0.6$, $\omega=21.0533\pi/T$;
(d) $A=0.7$, $\omega=15.8274\pi/T$.}
\end{figure}
\begin{figure}
\scalebox{0.4}{\includegraphics[scale=1]{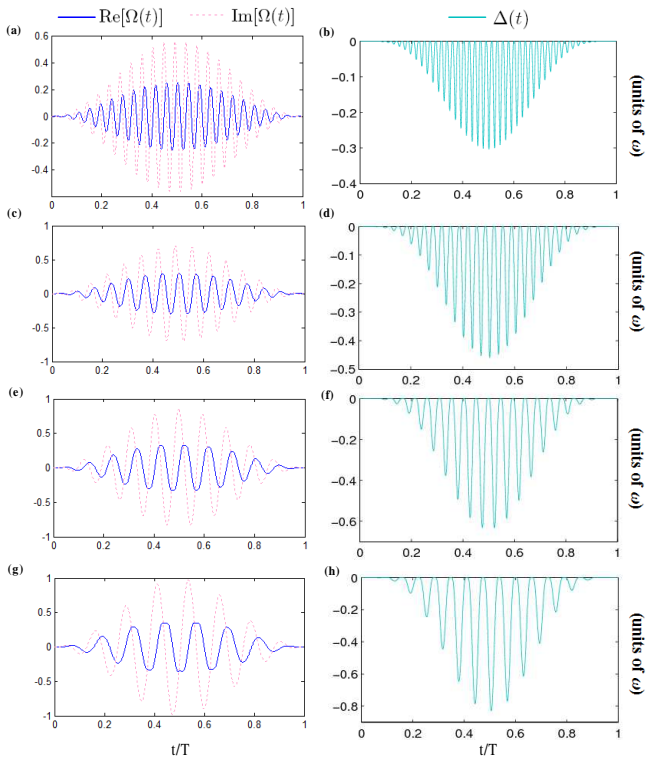}} \caption{The
real part $\mathrm{Re}[\Omega(t)]$ (the solid blue line) and the
imaginary part $\mathrm{Im}[\Omega(t)]$ (the dotted pink line) of
$\Omega(t)$, and the detuning $\Delta(t)$ (the light blue line)
versus $t/T$ with different parameters: (a)-(b) $A=0.4$,
$\omega=45.7220\pi/T$; (c)-(d) $A=0.5$, $\omega=29.7323\pi/T$;
(e)-(f) $A=0.6$, $\omega=21.0533\pi/T$; (g)-(h) $A=0.7$,
$\omega=15.8274\pi/T$.}
\end{figure}

According to Fig. 2, population transfers could be achieved with
$A=0.4,0.5,0.6,0.7$. This proves the invariant given in
Eq.~(\ref{e2}) is correct, and the parameters designed in this
section are valid. Moreover, it is easy to find out that the maximal
population of the intermediate state $|2\rangle$ increases slightly
when $A$ increases.

According to Fig. 3, the oscillations of pulses reduce when $A$
increases, since $A$ with larger value make the evolution of the
system go through fewer pulse periods (as shown in Table I). When
detunings and Rabi frequencies have too many oscillations, they may
be difficult to be realized in experiments. To reduce the
oscillations, one may increase $A$. However, from Fig. 3, the
maximal values of ratios $\Omega(t)/\omega$ and $\Delta(t)/\omega$
increase when $A$ increases. When $A$ is too large, the pulses and
detunings may go beyond the acceptable ranges. Therefore, when
designing pulses with smooth functions for a real experiment, one
should choose a suitable $A$ to make the pulses and detunings in
acceptable ranges.

\subsection{Pulse design with modifications around singular points}

In this part, we try to reduce the oscillations by choosing a flat
varying $\beta$. And we try to avoid the singularity of pulses by
modifying the pulses around their singular points. The modifications
are based on the fact that the Cauchy principal value of
\begin{equation}\label{b1}
\int_{x_0-\varsigma}^{x_0+\varsigma}\frac{1}{\cos x}dx
\end{equation}
is zero, where $x_0$ denotes a singular point of function $1/\cos
x$, and $\varsigma$ is an arbitrary small value. Therefore,
populations vary little in the time intervals around the singular
points of $1/\cos(\omega t)$. We can make some modifications of
pulses around the singular points of $1/\cos(\omega t)$.

Suppose that a population transfer starts at $t=0$ and ends at
$t=T$, and the initial state of the system is
$|\psi(0)\rangle=|1\rangle$. We maintain the condition
$\alpha=\pi/4$ in this section. Instead of $\beta$ and $\dot{\beta}$
shown in part A, we choose $\bar{\beta}$ and $\dot{\bar{\beta}}$,
respectively, as follows:
\begin{eqnarray}\label{b2}
&\bar{\beta}&=\frac{B}{2}[1-\cos(\frac{2\pi t}{T})],\cr\cr
&\dot{\bar{\beta}}&=\frac{\pi B}{T}\sin(\frac{2\pi t}{T}),
\end{eqnarray}
where, $B$ is a time-independent coefficient controlling the maximal
value of $\bar{\beta}$. In this case, pulses could still be smoothly
turned on and turned off. But different from the $\beta$ and
$\dot{\beta}$ designed in part A, the parameters that we chose here
could not eliminate the singularity of pulses. However, when $B=A$,
we have
\begin{eqnarray}\label{b3}
\bar{\epsilon}(T)=\omega\int_{0}^{T}\sin^2\bar{\beta}
dt\geq\omega\int_{0}^{T}\sin^2\beta dt=\epsilon(T).
\end{eqnarray}
So in the case of $B=A$, the maximal value of $\beta$ is
approximately equal to that of $\bar{\beta}$, but the population
transfer could be completed faster as it goes through fewer pulse
periods. That means the oscillations of pulses decrease a lot
compared with the results of part A. Moreover, the singular points
of pulses that we need to deal with are not too many. These results
could also be got by comparing Table II with Table I.
\begin{center}
\centering{\bf Table II. $B$ with corresponding $\omega T$.}
{\small\begin{tabular}{cc} \hline\hline
\ \ \ \ \ \ \ \ \ \ \ $B$\ \ \ \ \ \ \ \ \ \ \ \ \ \ \ \ \ \ \ \ \ \ &\ \ \ \ \ \ \ \ \ \ \ \ \ \ \ \ \ \ \ \ \ \ \ \ \ $\omega T$\ \ \ \ \ \ \ \ \\
\hline
\ \ \ \ \ \ \ \ \ \ \ $0.4$\ \ \ \ \ \ \ \ \ \ \ \ \ \ \ \ \ \ \ \ \ \ &\ \ \ \ \ \ \ \ \ \ \ \ \ \ \ \ \ \ \ \ \ \ \ \ $17.33\pi$\ \ \ \ \ \ \ \ \ \\
\ \ \ \ \ \ \ \ \ \ \ $0.5$\ \ \ \ \ \ \ \ \ \ \ \ \ \ \ \ \ \ \ \ \ \ &\ \ \ \ \ \ \ \ \ \ \ \ \ \ \ \ \ \ \ \ \ \ \ \ $11.34\pi$\ \ \ \ \ \ \ \ \ \\
\ \ \ \ \ \ \ \ \ \ \ $0.6$\ \ \ \ \ \ \ \ \ \ \ \ \ \ \ \ \ \ \ \ \ \ &\ \ \ \ \ \ \ \ \ \ \ \ \ \ \ \ \ \ \ \ \ \ \ \ $8.09\pi$\ \ \ \ \ \ \ \ \ \\
\ \ \ \ \ \ \ \ \ \ \ $0.7$\ \ \ \ \ \ \ \ \ \ \ \ \ \ \ \ \ \ \ \ \ \ &\ \ \ \ \ \ \ \ \ \ \ \ \ \ \ \ \ \ \ \ \ \ \ \ $6.13\pi$\ \ \ \ \ \ \ \ \ \\
\hline \hline
\end{tabular}}
\end{center}

Now, let us show how to deal with the singular points of pulses by
modifying the pulses around them. We take $B=0.5$ as an example. In
this case, a population transfer goes through more than 5 pulse
periods but fewer than 6 pulse periods. Since $\alpha=\pi/4$ is set,
we have
\begin{eqnarray}\label{b4}
\bar{\Omega}_p(t)=\bar{\Omega}_s(t)=\bar{\Omega}(t)=-\frac{1}{2\sqrt{2}\cos(\omega
t)}(2i\dot{\bar{\beta}}+\omega\sin2\bar{\beta}),\cr\cr
\end{eqnarray}
where, $\bar{\Omega}_p(t)$ and $\bar{\Omega}_s(t)$ are respectively
the pump and Stokes pulses decided by $\bar{\beta}$. There are
eleven singular points of $\bar{\Omega}(t)$ in this case. They are
$t_n=\frac{(2n-1)\pi}{2\omega}$ $(n=1,2,3...,11)$. We modify
$\bar{\Omega}(t)$ around these eleven singular points by
$\tilde{\Omega}(t)$ as follows:
\begin{eqnarray}\label{b5}
\tilde{\Omega}(t)&=\begin{cases} \bar{\Omega}(t_n-\delta
t)\cr+\frac{\bar{\Omega}(t_n+\delta t)-\bar{\Omega}(t_n-\delta
t)}{2\delta t}(t-t_n+\delta t), &t\in\Xi_n, \cr \bar{\Omega}(t),
&others,
\end{cases}\cr\cr&&
\end{eqnarray}
where $\Xi_n=(t_n-\delta t,t_n+\delta t)$ is the modifying interval
around the singular point $t_n$, and $\delta t$ is a parameter which
controls the length of modifying intervals around singular points.

By using Eq.~(\ref{b5}), we perform numerical simulations with
$\delta t=0.01~T$ and $\delta t=0.005~T$. In Figs. 4 (a) and (c), we
plot populations $P_1$, $P_2$, and $P_3$ versus $t/T$ with $B=0.5$
in the cases of $\delta t=0.01~T$ and $\delta t=0.005~T$,
respectively. In Figs. 5 (a) and (b), we plot the real part
$\mathrm{Re}[\tilde{\Omega}(t)]$ and the imaginary part
$\mathrm{Im}[\tilde{\Omega}(t)]$ of $\tilde{\Omega}(t)$ versus $t/T$
with $B=0.5$ in the cases of $\delta t=0.01~T$ and $\delta t=0.005\
T$, respectively. The detuning $\Delta(t)$ which is independent of
$\delta t$, is plotted in Fig. 5 (c).

\begin{figure}
\scalebox{0.5}{\includegraphics[scale=1]{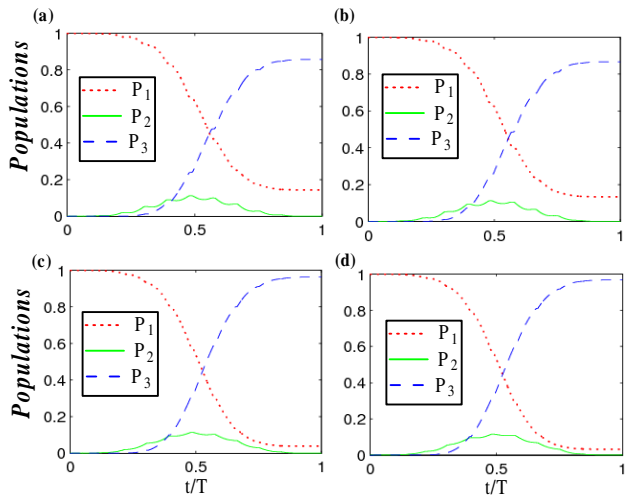}}
\caption{Populations $P_1$ (the dotted red line), $P_2$ (the solid
green line), and $P_3$ (the dashed blue) versus $t/T$ with $B=0.5$
in different cases: (a) $\delta t=0.01~T$; (b) $\delta t=0.01~T$
with $\mathrm{Im}[\tilde{\Omega}(t)]$ been neglected; (c) $\delta
t=0.005~T$; (d) $\delta t=0.005~T$ with
$\mathrm{Im}[\tilde{\Omega}(t)]$ been neglected.}
\end{figure}
\begin{figure}
\scalebox{0.52}{\includegraphics[scale=1]{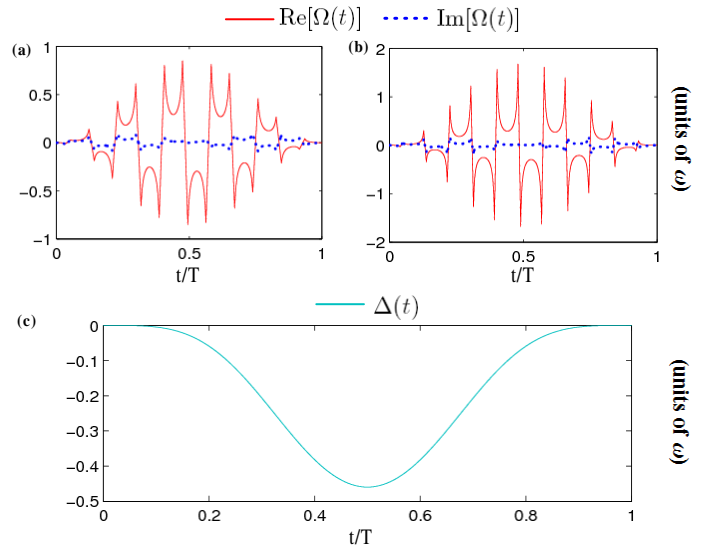}}
\caption{(a)-(b) The real part $\mathrm{Re}[\tilde{\Omega}(t)]$ (the
solid red line) and the imaginary part
$\mathrm{Im}[\tilde{\Omega}(t)]$ (the dotted blue line) of
$\tilde{\Omega}(t)$ versus $t/T$ with $B=0.5$ in different cases:
(a) $\delta t=0.01~T$; (b) $\delta t=0.005~T$. (c) The detuning
$\Delta(t)$ (the light blue line, independent of $\delta t$) versus
$t/T$ with $B=0.5$.}
\end{figure}

Seen from Figs. 4 (a) and (c), we find that population transfers are
imperfect, while the final population of $|3\rangle$ ($P_3(T)$)
increases when $\delta t$ reduces. For $\delta t=0.01~T$, we have
$P_3(T)=0.8516$, while for $\delta t=0.005~T$, we have
$P_3(T)=0.9618$. However, according to Figs. 5 (a) and (b),
increasing $P_3(T)$ by reducing $\delta t$ results in the increments
of amplitudes of pulses. In addition, we find that
$\mathrm{Im}[\tilde{\Omega}(t)]$ influences the population transfers
little for both $\delta t=0.01~T$ and $\delta t=0.005~T$, as
$\mathrm{Im}[\tilde{\Omega}(t)]\ll\mathrm{Re}[\tilde{\Omega}(t)]$.
We plot the population of each state versus $t/T$ with $\delta
t=0.01~T$ and $\delta t=0.005~T$ in Figs. 4 (b) and (d),
respectively, when $\mathrm{Im}[\tilde{\Omega}(t)]$ is neglected;
the numerical result shows $P_3(T)=0.8675$ ($P_3(T)=0.9680$) with
$\delta t=0.01~T$ ($\delta t=0.005~T$).

\subsection{Pulse design with reversely solved parameters}

In part B, we investigate pulse design with modifications around
singular points. The results show that the oscillations of pulses
could be reduced a lot. However, population transfers may be
imperfect if modifying intervals are not narrow enough. To decrease
the length of modifying intervals, we need to intensify the
amplitudes of pulses. Moreover, the forms of pulses may be complex
for the experimental realization. That motivates us to consider how
to design pulses with suitable forms, amplitudes and fewer
oscillations. In this part, we do not choose parameter $\beta$
directly, while we consider the Rabi frequencies of pulses first.

Let us start from Eq.~(\ref{e10}). Here, the condition
$\alpha=\pi/4$ is still adopted, such that
\begin{eqnarray}\label{c1}
\Omega_p(t)=\Omega_s(t)=\Omega(t)=-\frac{1}{2\sqrt{2}\cos(\omega
t)}(2i\dot{\beta}+\omega\sin2\beta).\cr\cr
\end{eqnarray}
Suppose that $\Omega(t)=\Omega_r(t)+i\Omega_i(t)$, Eq.~(\ref{e10})
can be replaced by
\begin{eqnarray}\label{c2}
&\Omega_r(t)\cos(\omega
t)&=-\frac{\omega}{2\sqrt{2}}\sin2\beta,\cr\cr
&\Omega_i(t)\cos(\omega t)&=-\frac{1}{\sqrt{2}}\dot{\beta},
\end{eqnarray}
where $\Omega_r(t)$ and $\Omega_i(t)$ are two real functions,
representing the real part and the imaginary part of $\Omega(t)$.
Parameters
\begin{eqnarray}\label{c3}
&\beta&=-\frac{1}{2}\arcsin[\frac{2\sqrt{2}}{\omega}\Omega_r(t)\cos(\omega
t)],\cr\cr
&\Omega_i(t)&=\frac{\dot{\Omega}_r(t)-\Omega_r(t)\omega\tan(\omega
t)}{\sqrt{\omega^2-8\Omega_r^2(t)\cos^2(\omega t)}},
\end{eqnarray}
can be solved from Eq.~(\ref{c2}). To make $\Omega_i(t)$ a bounded
function, it requires
\begin{eqnarray}\label{c4}
&\lim\limits_{t\rightarrow
t_m}[\dot{\Omega}_r(t)-\Omega_r(t)\omega\tan(\omega t)]&\cr\cr
&=\lim\limits_{t\rightarrow t_m}\frac{\dot{\Omega}_r(t)\cos(\omega
t)-\Omega_r(t)\omega\sin(\omega t)}{\cos(\omega t)}&=\mathrm{Const},
\end{eqnarray}
where, $t_m=(m+1/2)\pi$ ($m=0,1,2,...$). Furthermore, Eq.~(\ref{c4})
can be replaced by
\begin{eqnarray}\label{c5}
&&\lim\limits_{t\rightarrow t_m}[\dot{\Omega}_r(t)\cos(\omega
t)-\Omega_r(t)\omega\sin(\omega
t)]\cr\cr&&=\lim\limits_{t\rightarrow t_m}-\Omega_r(t)=0,
\end{eqnarray}
where, $\dot{\Omega}_r(t)$ is supposed to be a bounded function. It
means that, to avoid the singularity of $\Omega_i(t)$, we require
$\Omega_r(t_m)=0$.

Now, let us start from investigating pulses in a whole pulse period.
For example, the time interval $\pi/2\omega\leq t\leq5\pi/2\omega$
is considered. To fulfill the condition $\Omega_r(t_m)=0$, we simply
choose
\begin{eqnarray}\label{c6}
\Omega_r(t)=\Omega_0\cos^3(\omega t),
\end{eqnarray}
which is not difficult to be realized in experiments. It is easy to
obtain
\begin{eqnarray}\label{c7}
\Omega_i(t)=\frac{-4\Omega_0\cos^2(\omega t)\sin(\omega
t)}{\sqrt{1-\frac{8\Omega_r^2(t)}{\omega^2}\cos^2(\omega t)}}.
\end{eqnarray}
To avoid $\beta$ and $\Omega_i(t)$ taking complex values, it is
better to set $0\leq\Omega_0<\omega/2\sqrt{2}$. The increment
$\Delta\epsilon$ of $\epsilon$ in this pulse period can be
calculated by
\begin{eqnarray}\label{c8}
\Delta\epsilon=\omega\int_{\pi/2\omega}^{5\pi/2\omega}\sin^2\beta
dt,
\end{eqnarray}
via a numerical integration. We plot $\Delta\epsilon/\pi$ versus
$\Omega_0/\omega$ in Fig. 6. Moreover, $\Omega_i(t)/\omega$ versus
$t/(\pi/2\omega)$ and $\Omega_0/\omega$ are plotted in Fig. 7.
\begin{figure}
\scalebox{0.55}{\includegraphics[scale=1]{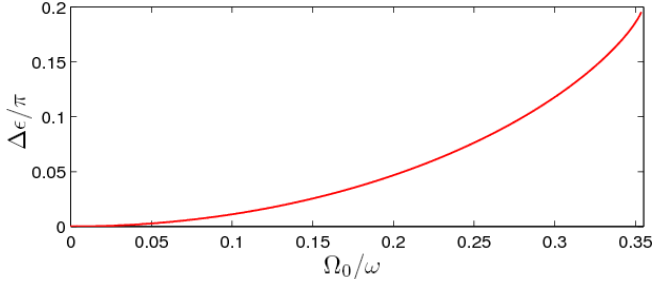}}
\caption{$\Delta\epsilon/\pi$ versus $\Omega_0/\omega$.}
\end{figure}
\begin{figure}
\scalebox{0.65}{\includegraphics[scale=1]{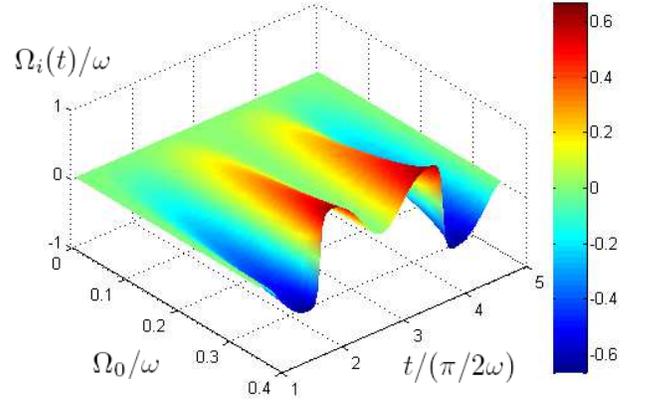}}
\caption{$\Omega_i(t)/\omega$ versus $t/(\pi/2\omega)$ and
$\Omega_0/\omega$.}
\end{figure}

From Fig. 6, to make the pulses have fewer oscillations, one can
increase the ratio $\Omega_0/\omega$ to make a population transfer
go through fewer pulse periods. On the other hand, according to Fig.
7, the amplitude of $\Omega_i(t)$ increases when $\Omega_0/\omega$
increases. For $\Omega_0$ that satisfies
$0\leq\Omega_0<\omega/2\sqrt{2}$, we have $|\Omega_i(t)|<0.64\omega$
($\forall t\in[\pi/2\omega,5\pi/2\omega]$). To make the operations
simple, $\Omega_0/\omega=0.3396$ is chosen, such that
$\Delta\epsilon=\pi/6$. Suppose that the parameters in every pulse
period repeat the results in $[\pi/2\omega,5\pi/2\omega]$. In this
case, if a population transfer starts at $t=\pi/2\omega$, it could
be finished at $t=25\pi/2\omega$, i.e., the population transfer goes
through 6 pulse periods.

The population of each state is plotted in Fig. 8. Furthermore, in
Figs. 9 (a), (b), and (c), $\Omega_r(t)$, $\Omega_i(t)$ and the
detuning $\Delta(t)$ versus $t/(\pi/2\omega)$ during the first pulse
period are plotted.
\begin{figure}
\scalebox{0.55}{\includegraphics[scale=1]{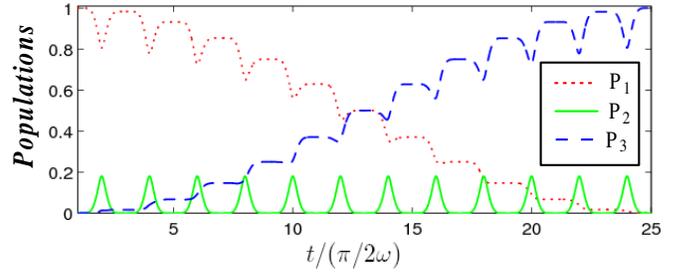}}
\caption{Populations $P_1$ (the dotted red line), $P_2$ (the solid
green line), and $P_3$ (the dashed blueline) versus
$t/(\pi/2\omega)$ with $\Omega_0/\omega=0.3396$.}
\end{figure}
\begin{figure}
\scalebox{0.48}{\includegraphics[scale=1]{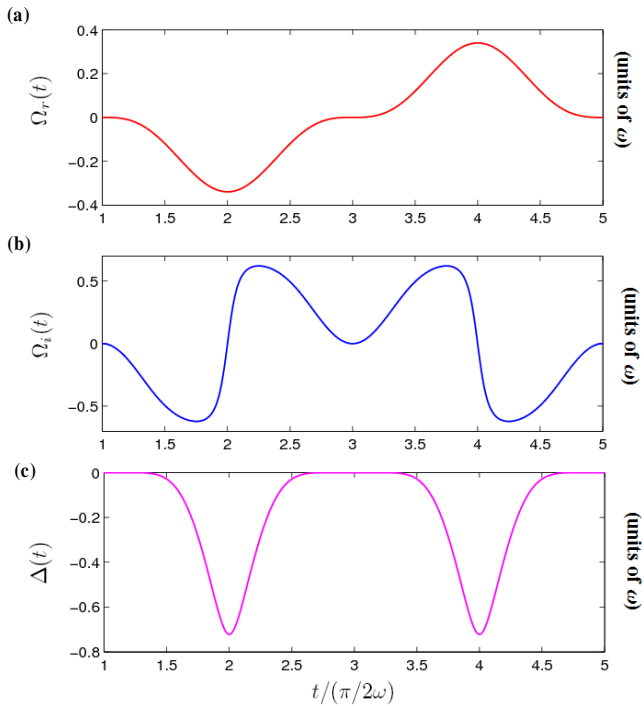}}\caption{(a)
$\Omega_r(t)$ versus $t/(\pi/2\omega)$; (b) $\Omega_i(t)$ versus
$t/(\pi/2\omega)$; (c) $\Delta(t)$ versus $t/(\pi/2\omega)$.
($\Omega_0/\omega=0.3396$)}
\end{figure}

As shown in Fig. 8, the population transfer can be achieved with the
designed pulses in this part. However, $P_3$ increases up to near
unity with greater and greater oscillations. The maximal hump of the
oscillations appears at $t=24\pi/2\omega$ with $P_3=0.806$.
Therefore, the real interaction time $T'$ should not approach
$24\pi/2\omega$ for a real experiment. To obtained $P_3\geq0.9999$,
we require $T'\geq24.71\pi/2\omega$ ($\delta T=|T'-T|\leq1.29\%$).
On the other hand, the pulses become weaker and weaker, and they
could be turned off smoothly at $t=25\pi/2\omega$. Besides, the
curve of $P_3$ has a platform at $t=25\pi/2\omega$. Therefore, $P_3$
still keeps near unity when $T'\geq25\pi/2\omega$. To summarize,
using the approach proposed in this part may have less robustness
against the operation errors of the interaction time compared with
the approaches of parts A and B. On the other hand, seen from Fig.
9, the real parts and the imaginary parts of pulses are much more
smooth compared with that of pulses which were designed in part B.
Moreover, since the population transfer goes through only 6 pulse
periods, the oscillations of pulses and detunings are much fewer
than that of pulses and detunings which were designed in part A.
Therefore, the approach of pulse design shown in this section may be
more attractive.

For the situation where the total interaction time $T$ is not the
integral multiple of a pulse period, e.g., $T=2p\pi/\omega+\tau$,
($p=0,1,2,...$), we can deal with the evolution of the system in the
$p$th pulse period by similar way for 1st pulse period. And then, we
only need to add pulse design for interval $[T-\tau,T]$ to make
population transfers successful at $t=T$.

\section{Conclusion}

In conclusion, we have proposed an invariant-based scheme for pulse
design without RWA. First, we found out an invariant for a
three-level system without RWA. Then, we exploited the invariant to
investigate pulse design for the population transfers in a
three-level system. From three different viewpoints, we gave three
approaches to design pulses in parts A, B and C of Sec. IV. In part
A of Sec. IV, we tried to design pulses with smooth functions. The
population transfers could be realized without singularity of
pulses. But the pulses would involve many oscillations. In part B of
Sec. IV, we tried to reduce the oscillations of pulses by modifying
the pulses around their singular points. The oscillations could be
reduced a lot, while the population transfers became imperfect and
the pulse forms might be complex for the experimental realization.
In part C of Sec. IV, instead of choosing control parameters
directly, we first chose pulses with feasible forms accompanied with
some undetermined coefficients. Then we reversely solved the control
parameters. With the help of numerical calculations, we determined
all the coefficients of pulses. With the approach shown in part C of
Sec. IV, feasible pulses could be designed for every pulse period,
and the oscillations of pulses could be well restricted.

Overall, the scheme has shown several novel results and advantages:

(i) To our knowledge, invariants for a three-level system without
RWA have not been investigated in the previous schemes. Therefore,
the invariant shown in Eq.~(\ref{e2}) may be a new one.

(ii) Based on pulse design with the invariant shown in
Eq.~(\ref{e2}), we do not need any extra couplings.

(iii) The amplitudes of pulses and the maximal values of detunings
could be well controlled in the present scheme. But it is difficult
for schemes with transitionless quantum driving to do so.

(iv) The pulses designed by the scheme can be smoothly turned on and
turned off. Therefore, the scheme should be robust against the
fluctuations of parameters.

(v) Compared with adiabatic processes, the system is not required to
satisfy the adiabatic condition, thus possessing higher evolution
speed.

With these advantages, the scheme may be useful for fast quantum
information processing without RWA.

\section*{Acknowledgement}

This work was supported by the National Natural Science Foundation
of China under Grants No. 11575045, No. 11374054 and No. 11674060,
and the Major State Basic Research Development Program of China
under Grant No. 2012CB921601.

\end{document}